# *How the Many Worlds Interpretation brings Common Sense to Paradoxical Quantum Experiments*

**Kelvin J. McQueen and Lev Vaidman**

*The many worlds interpretation of quantum mechanics (MWI) states that the world we live in is just one among many parallel worlds. It is widely believed that because of this commitment to parallel worlds, the MWI violates common sense. Some go so far as to reject the MWI on this basis. This is despite its myriad of advantages to physics (e.g. consistency with relativity theory, mathematical simplicity, realism, determinism, etc.). Here, we make the case that common sense in fact favors the MWI. We argue that causal explanations are commonsensical only when they are local causal explanations. We present several quantum mechanical experiments that seem to exhibit nonlocal "action at a distance". Under the assumption that only one world exists, these experiments seem immune to local causal explanation. However, we show that the MWI, by taking all worlds together, can provide local causal explanations of the experiments. The MWI therefore restores common sense to physical explanation.*

## 1. Introduction

The purpose of explanation is to help us understand why things happen as they do. Typically, one explains an event by citing the causes of that event. For example, to explain why the window broke, one might cite the throwing of a stone.

A particularly compelling and natural type of causal explanation is *local* causal explanation. In a local causal explanation there is spatiotemporal continuity between the cause and the explained effect. For example, the thrown stone explains the broken window in part because the stone traces out a continuous trajectory from the stone-thrower's hand to the window.

In opposition to local explanations are *nonlocal* explanations. Nonlocal explanations involve "action at a distance". Nonlocal explanations are rare in everyday life. Familiar cases are purported cases of psychokinesis, e.g., when people claim to bend spoons with their minds. If one explains the bending of a spoon in terms of an individual's mental effort, then one is offering a nonlocal explanation, since there is no spatiotemporally continuous series of causes and effects connecting the mental effort and the spoon bending.

Common sense rejects such nonlocal explanations and presumes a hidden local cause of the spoon bending. And there is a good reason for this. A nonlocal explanation of X in terms of Y only tells us *that* X happens given Y. But an explanation should enable us to understand *why* X happens given Y. This is what local causal explanations offer. We understand why the window broke given the stone-throw because we understand the spatiotemporal continuity between the cause and the effect. Thus, if any type of explanation deserves to be called "common sense explanation", it is local causal explanation. The point is a general one and is not specific to macroscopic objects. Whether we scale up to planets and galaxies or scale down to elementary particles, a causal explanation that leaves behind "gaps" in spacetime will in turn leave behind gaps in our understanding. Presumably this is why Einstein (1935) famously referred to action at a distance as "spooky".

In philosophy, there is an approach known as the *common sense tradition*. This approach takes various common sense beliefs as data for philosophical reflection and rejects those philosophical views that conflict with them (Lemos, ch.10, *this edition*). Here, we employ a similar approach in the context of physics. In particular, we argue that locality is a necessary condition for causal explanations to be commonsensical. We then take it as a requirement on an adequate physical theory that it be able to offer local causal explanations of all experimental outcomes.

In the next section, we discuss how the demand for local causal explanation shaped classical physics. In section 3, we introduce quantum physics in the context of a simple experiment, which, while strange, can be explained locally. We then specify three refined common sense principles of locality. These principles are satisfied by classical physics and by the simple quantum experiment. However, in sections 4 and 5, we present several quantum experiments that do seem to violate the principles and therefore defy common sense.

Finally, in section 6, we show that common sense can be restored to quantum theory if we adopt the many worlds interpretation (MWI). We show how the MWI is capable of giving local causal explanations of all these experiments. We conclude by discussing the significance of our results to debates over the interpretation of quantum mechanics. In particular, many reject the MWI by an appeal to common sense. As one philosopher puts it, "The most obvious disadvantage is that its burgeoning world of duplicate histories seem repugnant to common sense." (Papineau 1995, 239). We will argue that our results suggest the contrary, that the advantages to common sense that the MWI provides outweigh its disadvantages.

## *2. Common sense explanation in classical physics*

The philosopher of science W.V.O. Quine (1957, 229) once said, "Science is not a substitute for common sense but an extension of it". Accordingly, physicists seek local causal explanations of the motions of physical objects. This is clear in classical physics. For example, in Newtonian mechanics a body's motion in a given region and time is explained by a force applied at that region and time.

The exception is Newton's law of gravitation. According to this law, a planet's motion at a time is explained by a force applied by a *distant* planet at that time. However, Newton himself described this action at a distance as an "absurdity" and proclaimed the source of gravitational forces to be "hitherto unknown". In fact, Newton went so far as to say:

> "Tis inconceivable that inanimate brute matter should (without the mediation of something else which is not material) operate upon and affect other matter without mutual contact. […] [T]hat one body may act upon another at a distance through a vacuum, without the mediation of anything else, by and through which their action and force may be conveyed from one to another, is to me so great an absurdity that I believe no man who has in philosophical matters a competent faculty of thinking can ever fall into it." (Newton 2004, 102–3)

Newton's appeal to "a competent faculty of thinking" is an appeal to common sense. Newton is then proclaiming that his theory of gravity cannot offer any commonsensical explanation of gravitational motion, since it cannot offer a local causal explanation. (For recent debate over Newton's views on action at a distance see Ducheyne (2014).)

Later, Newton's Law was understood in local terms by adding the concept of a gravitational field created by massive bodies. Classical (i.e. non-quantum) physics, then, describes point particles ("point-masses") and fields spread out in spacetime. The point-masses create the fields which propagate with finite velocity. The motions of the point-masses are then affected locally by the fields of other point-masses. This physical ontology, together with Newton's laws, Maxwell's electromagnetic theory, and relativity theory, provides local explanations of many things we see around us.

The common sense picture of classical physics ultimately consists in the motions of particles on continuous trajectories and the continuous propagation of waves with multiple trajectories, which includes the local phenomena of (constructive and destructive) interference. The motion of particles and the propagation of waves provide (common sense) causal connections. This includes continuous traces that moving particles and propagating waves leave behind in their local environments.

The continuous propagation of waves and their local causal effects are illustrated by the Mach-Zehnder interferometer (MZI) (Zehnder 1891, and Mach 1892), see Fig.1. The electromagnetic wave (i.e. the photon) splits at the beamsplitter, passes through two arms of the interferometer and reunites at the second beamsplitter. The reuniting of the waves creates interference, which leads to observable effects. When the interferometer is properly tuned, only detector $D_2$ detects light. This is due to constructive interference towards $D_2$ and destructive interference towards $D_1$. If the distance of one path is lengthened by half of the wavelength of the electromagnetic wave, then only detector $D_1$ receives light.

Formally, classical physics places no constraints on interactions except that they be local. It is assumed that these local interactions never vanish completely and that the electromagnetic wave leaves some trace, e.g. heating of the air in the arms of the interferometer. Any such trace can in principle be detected, so the interferometer tuned as in Fig. 1a leaves an observable trace of the same form, see Fig. 1b.

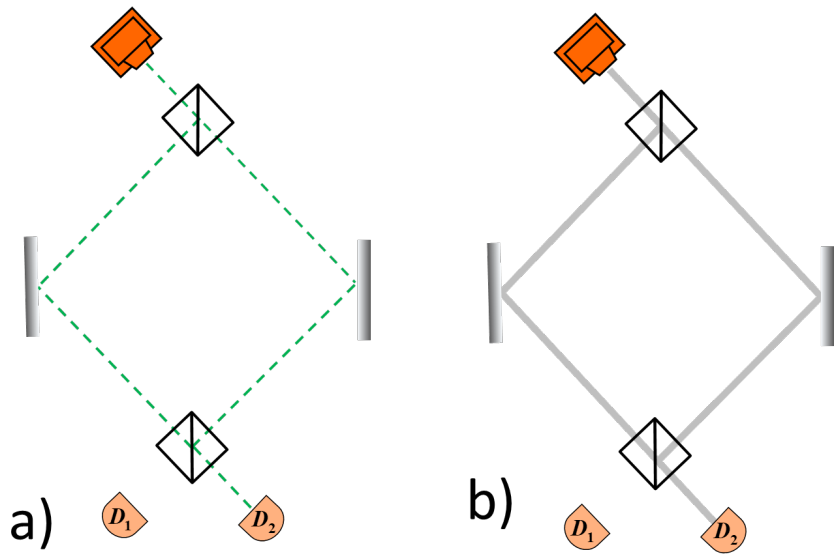

Figure 1: Mach-Zehnder interferometer (MZI). A beam of light is sent towards a beamsplitter, resulting in two beams of light. Each beam is then deflected towards a second beamsplitter, and then detected by one or both of the detectors. a) The interferometer is tuned so that the light comes only to detector $D_2$. b) The trace left by the light in the interferometer tuned as in a).

## *3. Common sense explanation in quantum physics*

In 1986 Grangier et al. showed that the MZI exhibits interference when we send single photons through the interferometer. MZI interference is also routinely observed with neutrons (Rauch 1974). But even before that, the double slit experiment with attenuated light, where single photons would one-by-one build up an interference pattern on the detection screen, led to the radical picture of quantum theory: In some sense particles are waves!

Let us first discuss a quantum particle and a single beamsplitter, Fig. 2. We observe that sometimes the particle passes undisturbed and sometimes it bounces from the beamsplitter as if the beamsplitter was a mirror. The difficulty is that in standard quantum theory there is no equation describing the motion of the particle to one direction *or* the other. The equation we have is the wave equation (the so-called Schrödinger equation), which splits the incoming wave to two waves, one reflected and one undisturbed, both propagating on a straight line, see Fig. 2a.

In quantum theory all couplings are local, and these local couplings never vanish. For example, when a particle passes by, it always leaves some trace in its surrounding environment. However, the trace, i.e. the local change of the quantum state of the environment, is not always detected in a single system. In fact, in a well-performed interference experiment, the probability of finding the change due to the passage of a single particle is very small. We can convince ourselves that there is such a change only by performing the experiment on a large ensemble of identically prepared particles. The trace shows the splitting of the wave, see Fig 2b.

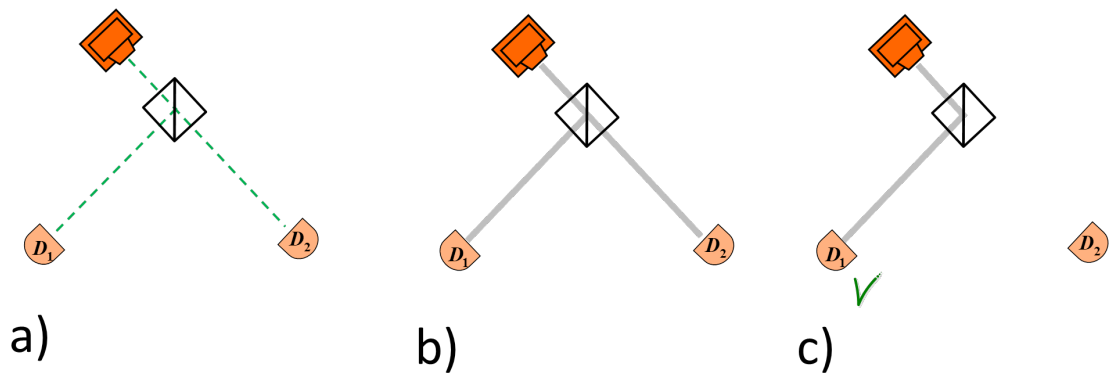

a)   b)   c)

Figure 2: A particle source, beam splitter, and two particle detectors. a) The quantum wave of the particle. b) The trace of particles sent through the beamsplitter, found by measuring the local environment of the two arms of the interferometer as a large ensemble of particles are sent through. c) The trace of a postselected ensemble of particles sent to the beamsplitter and detected by $D_1$. Particles that reach $D_1$ do not leave a trace on the path to $D_2$.

The trace we obtain can vary depending upon *postselection*. For example, the trace in Fig. 2c is obtained by examining only those particles that were found in $D_1$. But a different trace is found if we postselect only those particles found in $D_2$. In that case we would instead find a single continuous trace leading from the source to $D_2$.

Note that there is no classical wave analog for such behavior, since there is no meaning for a classical postselected wave. Only in quantum mechanics are future measurement outcomes not fully specified by a complete description of the initial state. The trace of a postselected particle in Fig. 2c is analogous to the trace of a classical particle reflected by the beamsplitter toward $D_1$.

In Fig. 2b there is no postselection and the trace corresponds to the quantum wave of each particle. This might conflict with common sense: for how can a single particle be in two places simultaneously? The lesson of quantum mechanics is that particles are waves and thus they can be simultaneously in several places. It is strange and unusual, but since classical waves are well understood, the wave behavior of a quantum particle can be accepted as sensible. We can get used to it just as we have gotten used to the idea that there is no observer-independent simultaneity of spatially remote events in special relativity.

Consider again the MZI in Fig. 1b, but this time with quantum particles instead of classical waves. We tune the MZI so that, due to constructive interference towards $D_2$, all particles are found in $D_2$. In that case, the trace is identical to the trace found in the MZI with classical waves, depicted in Fig. 1b. When we detune the interferometer in a way that spoils the destructive interference towards $D_1$, but postselect the particles detected in $D_2$, the trace is again as in the tuned interferometer, Fig. 1b. This is in contrast with the not tuned classical wave interferometer, which leaves a trace also on the path leading toward $D_1$.

Of some interest is a sequence of two MZI interferometers both tuned so that the input from the left constructively interferes towards the right output port. Quantum calculations show that the postselected particles in detector $D_1$ or $D_2$ leave a single path trace, see Fig. 3. The lack of a trace in the right (left) arm of the top MZI makes sense, since the tuning of the bottom MZI prevents the particle in the right (left) arm of the top MZI from reaching detector $D_2$ ($D_1$).

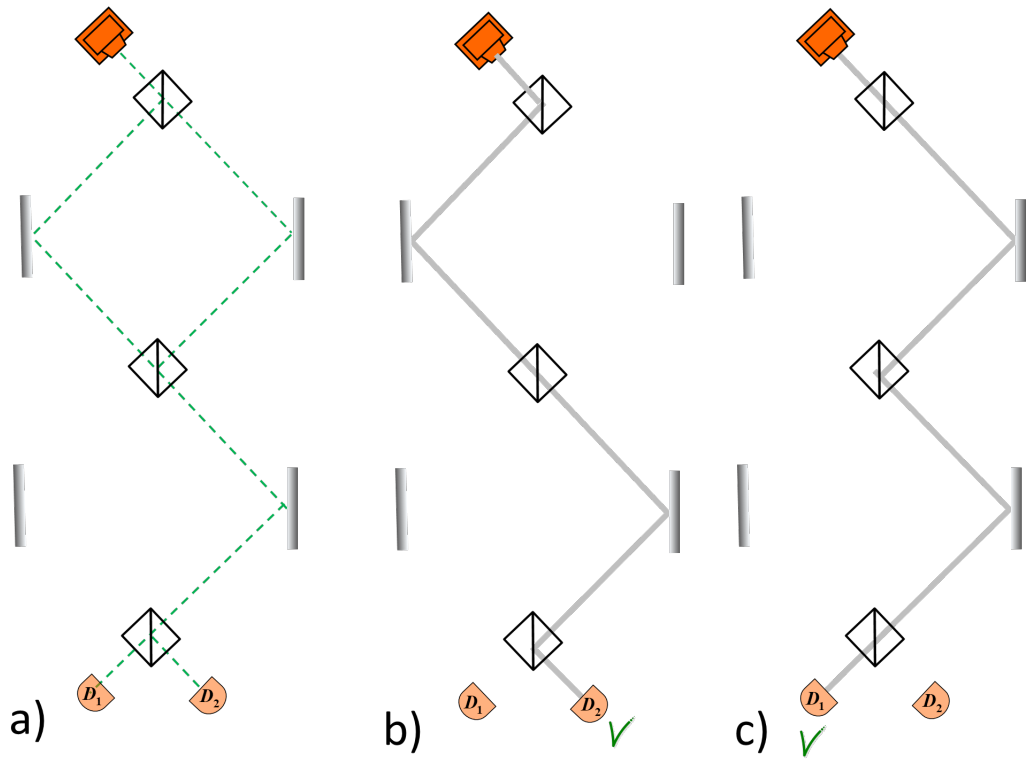

Figure 3: a) Quantum state of particle sent through a double MZI, each MZI is tuned as in Fig. 1. b,c) Traces of postselected particles detected in one of the detectors show a continuous path.

All of these cases of particles behaving like waves are strange, but they do not defy common sense explanation, understood as causal explanation in terms of continuous local trajectories/propagations through space.

Let us recap our arguments so far, for our claim that locality is a necessary condition for causal explanations to be commonsensical. *First*, the locality condition is supported by layman cases: the thrown stone is a common sense explanation of the window's breaking whereas mental effort is not a common sense explanation of spoon-bending. *Second*, it is possible to explain why the locality condition is commonsensical: non-local causal explanations do not perform a defining function of explanation: they only tell us *that* the effect occurs given the cause, without telling us *why* the effect occurs given the cause. *Third*, the locality condition is not restricted to layman cases, since the reason for its applying to layman cases (discussed in the second point) passes over naturally to any scale. *Fourth*, the scale-neutral generality of the condition is illustrated by the history of physics, which has consistently sought local explanations, gravity being a paradigmatic example. *Fifth*, it is widely accepted by physicists and philosophers of physics that the locality condition is commonsensical. We have illustrated this with quotes from Newton and Einstein, and it is not hard to find contemporaries describing the condition as "very commonsensical" (Lange 2002, 94). In what follows, we present interferometer experiments that seem to exhibit nonlocal action at a distance. If they intuitively violate common sense because they exhibit this feature, then they provide further support for the locality condition.

The paradoxical experiments discussed in the next two sections concern the wave nature of quantum particles, traces left behind by such particles, and information carried by such particles. It will therefore be useful to define more specific principles for each of these, which follow from the locality condition:

(i) Particles (which are waves in quantum mechanics) move on continuous trajectories (in some cases, *multiple* continuous trajectories).
(ii) A trace is left behind in a region if and only if a particle went through that region.
(iii) Information sent from one location to another always leaves a trace of some information carrier.

In the next section, we will present an example of an interferometer that apparently contradicts (i) and (ii). Then in section 5, we will present a more complex interferometer that apparently contradicts (iii).

## *4. The nested interferometer*

Let us now consider a MZI in which one arm is replaced by a smaller MZI, which is tuned to create constructive interference toward the final beam splitter of the external MZI, see Fig. 4. When we also tune the external interferometer to create constructive interference towards $D_2$, the situation seems very similar to a standard MZI. Indeed, the traces left behind are unsurprising as there is a set of continuous trajectories from the source to the detector.

The surprising situation contradicting common sense happens when we tune the inner interferometer to create destructive interference towards the final beamsplitter of the external interferometer, see Fig. 5. Now the external interferometer is not an interferometer any more. There is only one possible trajectory from the source to the detector. If we get the click at $D_1$, common sense tells us that the particle had a well-defined trajectory down arm $C$. The surprising feature is the trace left by the particle. It is not only on path $C$. Traces of similar strength also appear in the inner arms of the nested interferometer, see Fig. 5b. The particles leave traces in places through which they could not pass!

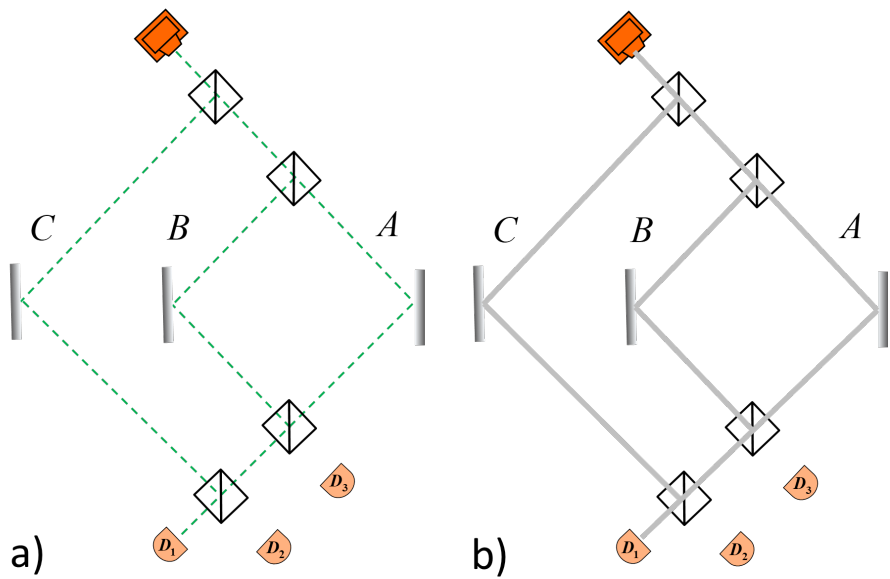

Figure 4: a) Nested MZI tuned to constructive interference towards $D_1$. b) Traces of particles correspond to the locations with nonvanishing wave function.

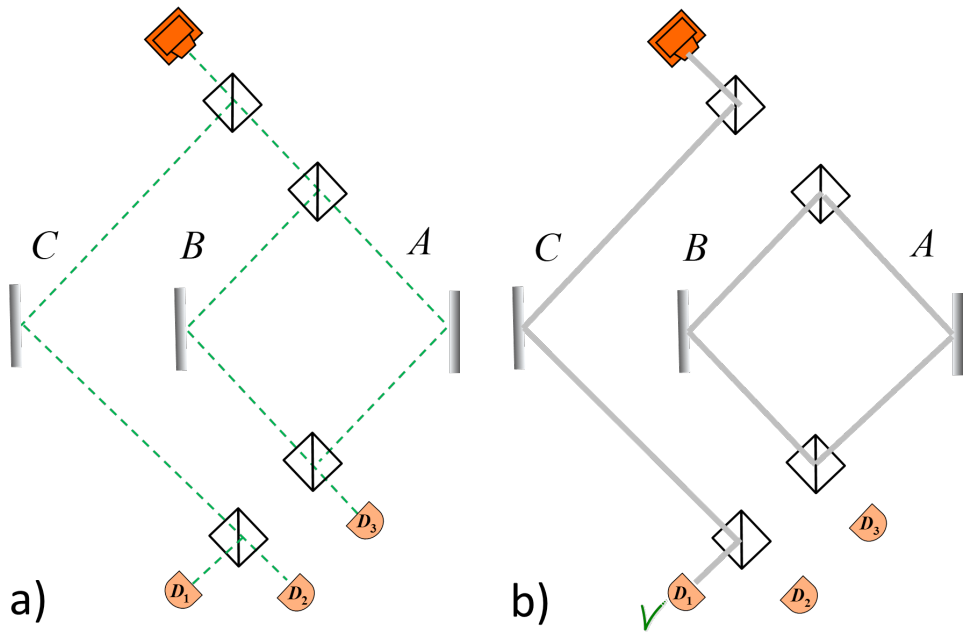

Figure 5. a) Nested MZI with inner interferometer tuned to deststructive interference towards the second beamsplitter of the large interferometer. b) Traces of particles detected at $D_1$ appear on path $C$ but also, surprisingly, inside the inner interferometer.

There is a large controversy about the nested MZI, which began with the analysis in Vaidman (2013). Experiments were done with continuous light (Danan et al. 2013), with single photons (Zhou 2017), and even with neutrons (Geppert-Kleinrath 2018) showing (although somewhat indirectly) the trace in the isolated region. However, there is no agreement about the meaning of these readings (Vaidman 2018) and the loudest criticism was by Englert et al. (2017) in a paper originally titled "Past of a quantum particle: Common sense prevails". See the reply and multiple references to other discussions in Peleg and Vaidman (2019). Many aspects of the experiment were criticized. The main line of criticism was that the existence of the trace of the particle inside the inner interferometer invariably spoils the coherence of the photon wave function and thus invariably spoils exact destructive interference. So, it is not true that there is precisely zero trace on the way towards and on the way out of the inner interferometer. This, however, hardly resolves the problem. We can improve the fidelity of the interferometer by reducing the local coupling of the particle and its environment. This makes the trace outside the inner interferometer arbitrarily small relative to the trace inside the interferometer, which is of the same order of magnitude as in the arm $C$ where everyone agrees that the particle was present. Since the interaction of the particle with the environment is the same inside and outside the interferometer, there is no explanation of the drastic difference in the trace.

To summarize. Common sense physics requires that (i) particles (waves) move on continuous (and sometimes multiple) trajectories; and that (ii) a trace is left behind in a region if and only if a particle went through that region. The trace they leave is due to local interactions with the environment in the locations where they were present. The strength of the trace is a function of the interaction and the time the particle spent in a particular location. In the nested MZI, the trace on the continuous trajectory *C* is commonsensical, since the particle can follow this (and only this) trajectory. However, there is also a trace of the same strength in the inner box, which has no common sense explanation given principles (i) and (ii).

Vaidman (2013) defined the places with significant traces as places where the particle was. But we do not need this definition to obtain a contradiction with common sense: a significant trace in a location without a trace leading to that location is enough to contradict the common sense principles. This contradiction will be resolved, with the help of many worlds, in section 6. But first, let us consider one more paradoxical quantum experiment.

## *5. Counterfactual communication*

Information (signals) cannot be sent faster than light. Common sense tells us that information can be transferred from one place to another with some carrier of information, particles. Since no particle can move faster than light, no signaling can happen faster than light. It seems obvious that to send information from one place to another we need to send particles between these two places, e.g. photons.

However, as impossible as it might sound, quantum mechanics allows us to send signals, to communicate, without particles in the transmission channel. We do need a transmission channel. In the communication protocol, sometimes (very rarely) the particles do pass through the channel, but these are (rare) events of communication failure. These events are discarded. The communication takes place only when particles were not in the transmission channel. It is enough that they could have been there: this is why these protocols are named "counterfactual" communication protocols.

Counterfactual communication protocols were inspired by interaction-free measurements (IFM), (Elitzur and Vaidman 1993) and were first proposed and performed by Hosten et al. (2006) with the name "counterfactual computation". They were rediscovered in 2013 (Salih et al. 2013) and were later upgraded to communicate quantum information (Salih 2016 and Li et al. 2015).

The protocols are based on a complicated interferometer. Consider Alice and Bob, who are trying to communicate with each other. A major part of the interferometer is in Alice's site and some part of the interferometer is in Bob's site, see Fig. 6. The sites are separate and between the sites there are numerous transmission channels. Alice sends a single photon into the interferometer. Bob sends a signal to Alice by blocking or not blocking the arms of his part of the interferometer. If he does not put the block up, the photon ends up in one port at Alice's site, and if he blocks the arms, the photon ends up in another port of Alice.

In rare cases, the photon is absorbed at Bob's site, but those cases are discarded. In cases where Alice detects the photon (probability close to 1), the information is sent from Bob to Alice.

What is the argument that the protocols are counterfactual? Standard quantum theory does not have a definition of when the particle is present in the channel. Except for cases of single localized wave packets, when it can be considered similarly to a classical particle (or in some situations in the Bohmian interpretation (Bohm 1952), we do not know through which arm it passed in the MZI. This is similar to not knowing which slit the particle passes through in the famous double slit interference experiment. The counterfactuality of the above protocols was based on a classical physics argument: The photon was not in the transmission channel during the communication because if it were there, it could not end up in one of Alice's detectors. This can be seen in Fig. 6. If Bob does not put the block up, any photon that passes through the transmission channel will not reach the next beamsplitter that would take it to Alice (Fig. 6a). If Bob does put the block up, then any photon that passes through the transmission channel is simply absorbed by his block.

One of us (Vaidman 2007, 2014, 2015, 2016) criticized this counterfactuality argument on the basis that the particle in these protocols leaves a trace in the transmission channel, which is even stronger than the trace of a particle actually passing through this channel. This is the same type of trace as described in the previous section. The postselected particle could not pass through the inner interferometer, but it left a trace just as a particle actually present there would leave.

The full counterfactual communication protocols which can reliably and efficiently transmit both classical bits are fairly complicated, see Fig. 6. One of their crucial ingredients is the quantum Zeno effect, which requires numerous (counterfactual) passages of the photon between the parties. We refer the reader to the original papers for their descriptions. Here, following Aharonov and Vaidman (2019), we will only explain how to correct a simple protocol, which will be capable of transmitting in a counterfactual manner (without a trace of the particles in the transmission channel), the bit 0. The modification which makes full (i.e. transmitting both bit 0 and bit 1) counterfactual protocols counterfactual is essentially the same.

The description of the protocols that separately transmit bits 0 and 1 in a counterfactual way will be enough to demonstrate the conflict between common sense and counterfactual communication. Let us start with interaction free measurement (IFM) (Elitzur and Vaidman 1993). The interferometer is tuned like the MZI from before, see Fig. 7a. To send bit 1, Bob blocks his arm of the interferometer. Now detector $D_1$ has probability 25% to click. When the click happens, Alice knows that Bob blocked the arm of the interferometer, i.e., he transmitted bit 1. When the block is present, there is no trace in the right arm of the interferometer. Since there is no trace in the transmission channel, we can claim that the photon was not there and therefore it is counterfactual communication.

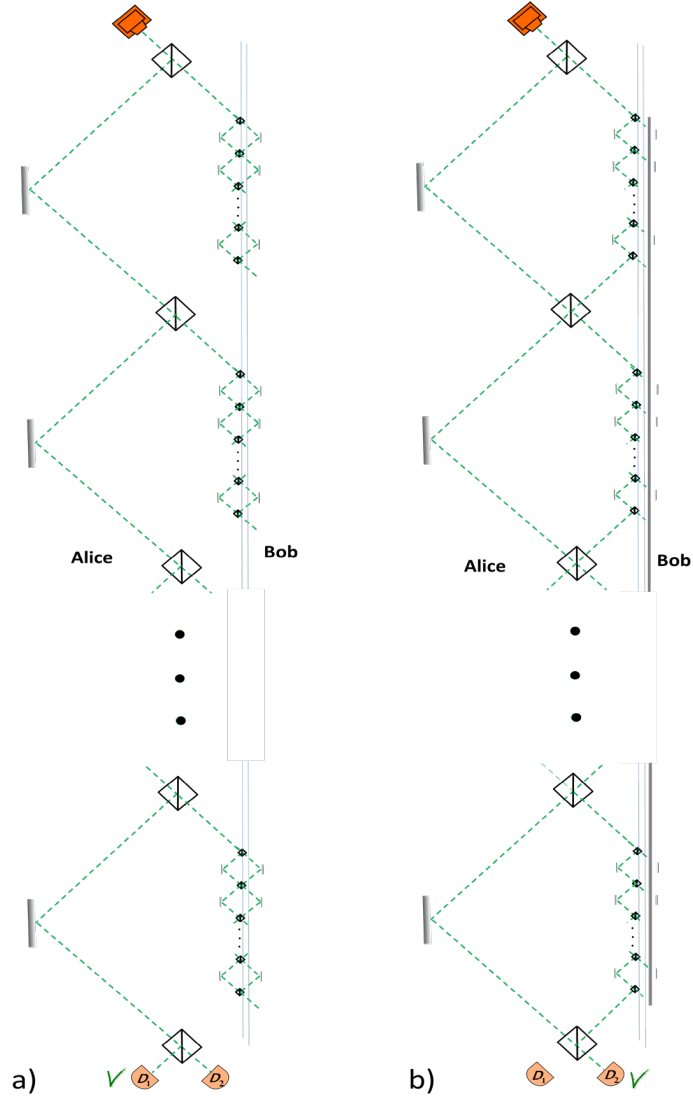

Figure 6. Counterfactual communication protocols. The interferometer consisting of $N$ external interferometers each having a chain of $M \gg N$ inner interferometers is mostly in Alice's site separated from Bob's site by two planes. Between the planes, there are numerous paths that photons might go through. In the protocol a single photon is sent by Alice into the interferometer. a) To send bit 0, Bob does nothing. In this case detector $D_1$ clicks with probability close to 1. With probability of order $1/N^2$ there might be an error and $D_2$ clicks instead. With probability of order $1/N$ the photon passes through the transmission channel to Bob's site and is not detected by any of Alice's detectors. b) To send bit 1, Bob blocks all arms of the interferometer in his site. In this case, detector $D_2$ clicks with probability close to 1. With probability of order $N/M$ the photon passes through the transmission channel to Bob's site and is not detected by any of the Alice's detectors. Given the criterion of the past of the particle, which states that the particle was where it left a trace (Vaidman 2013), all full counterfactual communication protocols that have so far been proposed are actually not counterfactual. We can still argue that the counterfactual communication of even one of the bit values contradicts common sense, but there is no need to weaken the claim. Recently a simple modification of full communication protocols was proposed. It makes all protocols counterfactual (Aharonov and Vaidman 2019).

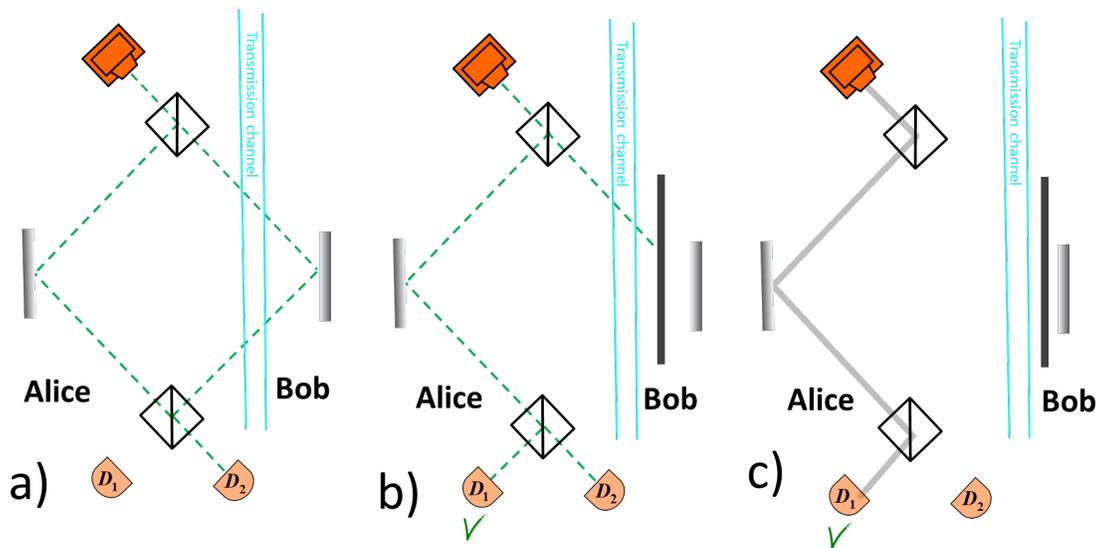

Figure 7. Counterfactual transmission of bit 1. a) MZI tuned to destructive interference towards $D_1$. b) When Bob blocks his arm of the interferometer detector $D_1$ can click. This click is the transmission of bit 1 from Bob to Alice. c) When the photon is detected at $D_1$ no trace is present in the transmission channel, which makes this communication counterfactual.

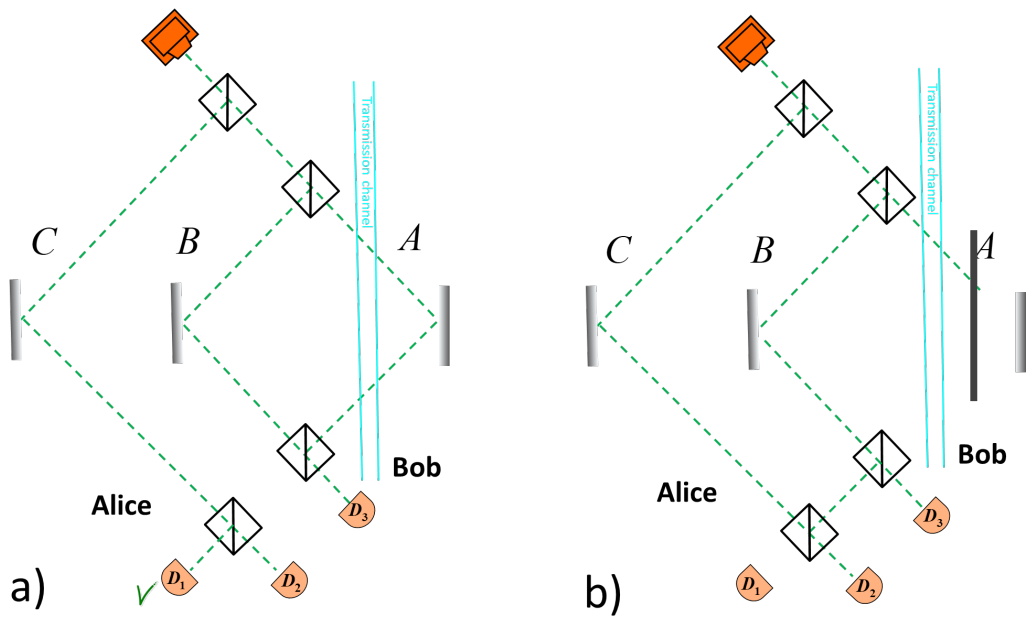

Figure 8. a) Naïve counterfactual transmission of bit 0. The nested MZI is tuned with two requirements: a) Without Bob's block, there is destructive interference of the inner interferometer towards the beamsplitter of the external interferometer. b) When Bob blocks his arm of the interferometer, there is destructive interference towards detector $D_1$.

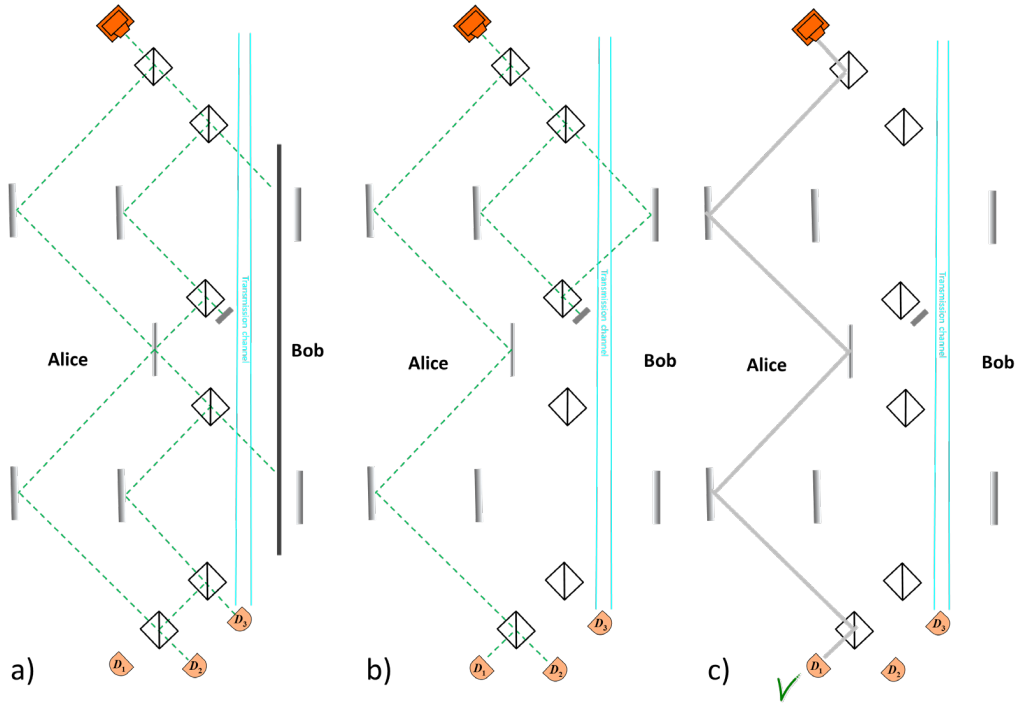

Figure 9. Counterfactual transmission of bit 0. Two nested interferometers connected by a double-sided mirror. (a-b) shows the tuning of the interferometer with and without the block of Bob. c) The trace in the interferometer when Bob does not block his arm. $D_1$ can click only when there is no block and then there is no trace in the transmission channel.

The next ingredient of the counterfactual protocols is the transmission of bit 0. Naively, it can be achieved using a properly tuned nested MZI, see Fig. 8. On the one hand, it is tuned as in the interferometer in Fig. 5a, such that the particle from the inner interferometer cannot reach detectors of the external interferometer, see Fig. 8a. On the other hand, when the inner interferometer is blocked we arrange destructive interference towards $D_1$. Now, the click in $D_1$ tells us that Bob did not use the block, i.e., the value of the bit is 0. Naively it looks counterfactual, since the photon in the arm of the external interferometer that enters the inner interferometer cannot reach $D_1$. But, as can be seen from Fig. 5b, the photon leaves a trace inside the inner interferometer and thus it leaves a trace in the transmission channel.

Fortunately, there is a simple modification of the protocol which makes it counterfactual (Aharonov and Vaidman, 2019). Two nested interferometers connected by a double-sided mirror with proper tuning achieves this task, see Fig. 9. Detector $D_1$ cannot click if Bob uses the block, so the click in the detector tells Alice that there is no block, that the bit is 0. And in this case there is no trace in the transmission channel. For an explanation for the absence of the trace, see Aharonov and Vaidman (2019).

In Fig. 7 we showed a counterfactual communication of bit 1 and in Fig. 8a counterfactual communication of bit 0. This is not enough for two reasons. First, both protocols have a large probability to fail. Second, these are different devices, so we cannot really send different bits in a counterfactual manner, each bit requires a different device for counterfactual transmission. The Quantum Zeno effect allows us to combine the two elements in a complicated interferometer shown in Fig. 6, to resolve the two problems together. The original solution falls short, because it had a trace when bit 0 was communicated. But a simple modification shown in Fig. 9. corrects this problem. Just double each external interferometer and connect it by a double-sided mirror (with appropriate tuning of the interferometer). (Note also a very recent simple reliable full counterfactual protocol with this modification (Vaidman 2019) which, however, is highly inefficient, since 95% of the sent photons are lost.)

The details can be seen in the physics references. The bottom line is that a full communication channel can be achieved with no trace left by the information carriers between Alice and Bob. And it is not by building a super-technology fiber in which photons leave no trace. The physics of the transmission channel is such that a photon, when present, would leave a trace. Still, in the communication, no trace is left. There is only a small probability for failure: then a trace is left (and no transmission happens). We therefore have a contradiction with our third principle of common sense physics, according to which information sent from one location to another always leaves a trace of some information carrier.

## 6. Common sense regained: the many worlds interpretation

The phenomena contradicting common sense are the phenomena violating our principles (i)-(iii) from section 3. This includes a trace left in a location without a trace that led to that location and information sent from one location to another without a trace of any information carriers.

Such phenomena occur even when the environment is not a special one: if particles are certain to be in these locations, they invariably leave a trace. We know that these phenomena are caused by particles that are sent through the experimental setups, since nothing happens when such particles are absent. In our world, we observe these phenomena. Therefore, in our world, common sense does not prevail.

According to the many worlds interpretation of quantum mechanics (MWI) (Everett 1957; Vaidman 2014b), quantum experiments with several possible outcomes give rise to newly created worlds corresponding to those outcomes. These worlds are not fundamental objects in the theory: they are emergent patterns that help to explain our experience. Experiences of distinct outcomes after a given measurement correspond to different worlds. These worlds have mathematical counterparts in the quantum wave function of the universe, which correspond to well-localized macroscopic objects, such as measuring devices, people, planets, etc.

In the MWI, the laws of physics (in particular, the Schrödinger wave equation) describe all worlds together, and do so in a way that restores common sense principles (i)-(iii). The MWI resolves the contradictions with common sense in the examples described above, while retaining their surprising character. The MWI resolution is not that the examples somehow lose their paradoxical features *in our world* or in parallel worlds. Instead, common sense principles (i)-(iii) are restored in the physical universe that incorporates all worlds together.

Here is the MWI explanation of the nested interferometer. In our world, with detection of the photon by detector $D_1$, there is a paradoxical trace in a separate island with no trace leading to it, Fig. 5b. However, we should understand that there are also two other parallel worlds where the photon is detected at other detectors, see Fig. 10a and Fig. 10b. The world with detection at $D_2$ also has a paradoxical trace, Fig. 10a, while the world with detection at $D_3$, has a continuous trace, Fig. 10b. Since physics describes all worlds together, the requirement of continuous trace is for all worlds together. The traces in all worlds together, i.e. the places where the environment changed its quantum state in the world with a click in $D_1$, in the world with a click in $D_2$, and in the world with a click in $D_3$ are shown on Fig. 10c. There is no isolated island of trace created by the particle. Traces are created by particles (waves) moving on (multiple) continuous trajectories. Therefore, there is no contradiction with common sense principles (i)-(iii).

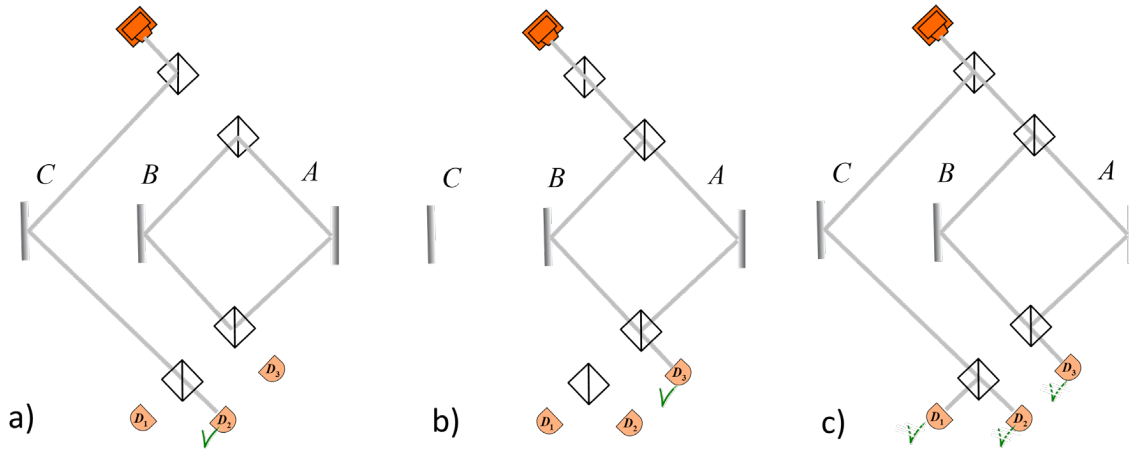

Figure 10. a) The trace in the world with the click at detector $D_2$. b) The trace in the world with the click at detector $D_3$. c) The trace in the physical universe, which includes all worlds together (Fig. 5b. and Figs. 10a,b.) No separate islands of trace, no contradiction with common sense.

Let us now explain how the MWI brings common sense to counterfactual communication protocols. In the IFM in the presence of the block, i.e. counterfactual communication of bit 1, presented in Fig. 7, we find in our world that Bob's arm of the interferometer is blocked without any trace near the block. In Fig. 11, the traces in parallel worlds of this experiment are shown. Fig. 11a describes traces in a world without any trace near the block, but the world is not problematic, since the click in $D_2$ provides no information about the presence of the block. Fig. 11b shows the trace when the photon is absorbed by the block and Fig. 11c presents the traces in all worlds together. In the full physical universe, there is a continuous trace towards the block and it is clear that it stops the photon from going towards Alice.

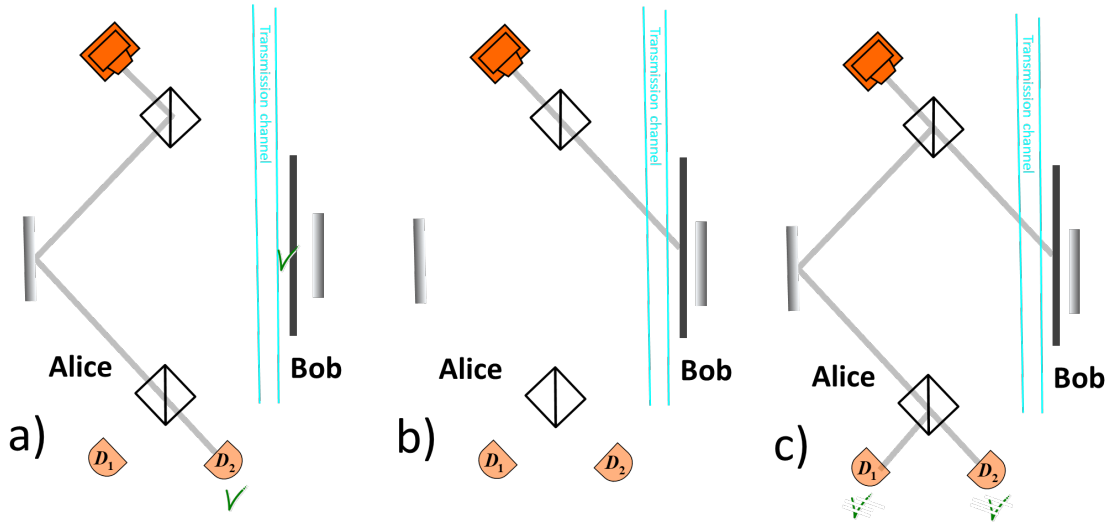

Figure 11. a) The trace in the world with the click at detector $D_2$. b) The trace in the world without detector clicks in which the photon is absorbed by the block. c) The trace in the physical universe, which includes all worlds together (Fig. 7c and Figs. 11a,b.). The photon left a trace in the transmission channel.

In the counterfactual communication of bit 0, presented in Fig. 9, we find that there is no block in Bob's arms without any trace at Bob's site. In the parallel world with the click of detector $D_2$, see Fig. 12a, we also do not have any trace at Bob's site. However, there is also a world without clicks of Alice's detectors, Fig. 12b, in which the photon passes through one of Bob's arms leaving a trace there. Fig. 12c shows traces in all worlds together. The physical laws applied to the physical universe incorporating all worlds together describe this experiment in terms of common sense continuous trajectories and local causality. The example is still very surprising, because we need ingenuity to construct experiments whose results provide useful information in a particular world due to interaction in parallel worlds.

In the full counterfactual communication protocol, the explanation is the same. We successfully communicate in our world because in parallel worlds the photons travel in the transmission channel between Alice and Bob and leave a trace there. What is even more surprising is that postselection here is almost not needed: with very high probability, the protocols work, and only very rarely does the photon fail to be detected by Alice. Although there is a very large number of these failure worlds with a particle in the transmission channel, we, nevertheless, have only a tiny probability to find ourselves in them (the meaning of self-location probability in the MWI is discussed in McQueen and Vaidman (2019)). This is the paradoxical quantum Zeno effect. Still, the causality issue is resolved: in all worlds together the particle passes (leaves a trace) through the transmission channel.

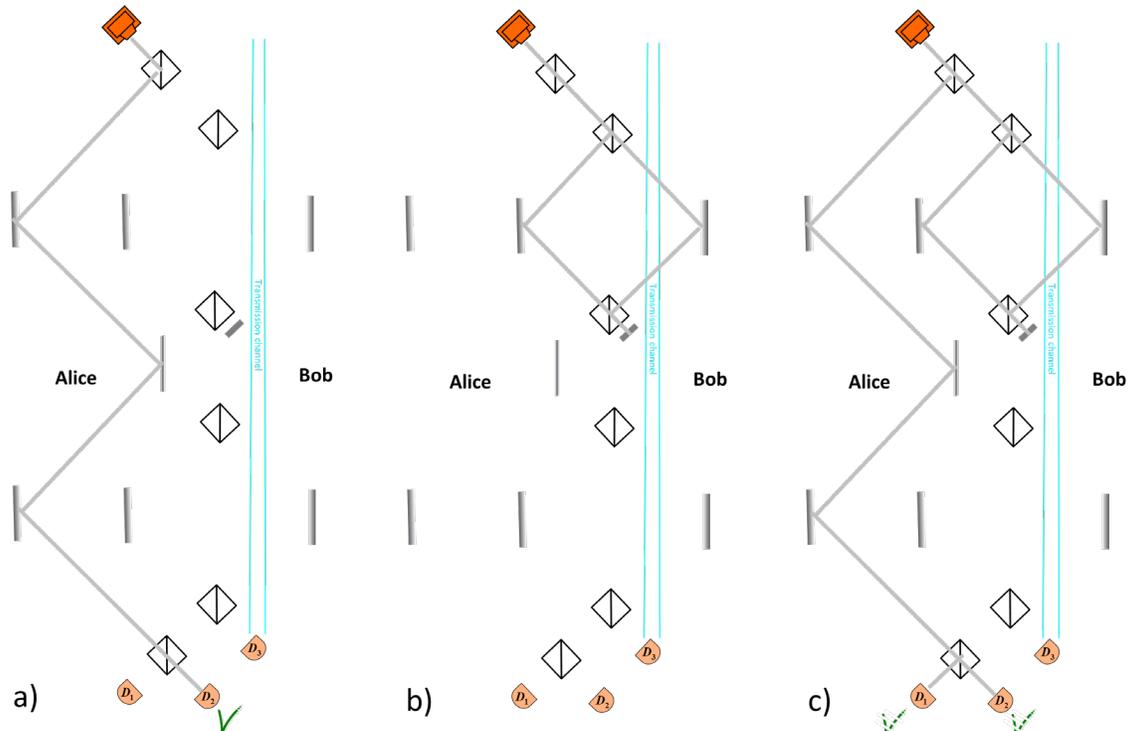

Figure 12. a) The trace in the world with the click at detector $D_2$. b) The trace in the world with the click at detector $D_3$. c) The trace in the physical universe, which includes all worlds together (Fig. 9c and Figs. 12a,b.). The photon left a trace in the transmission channel.

To summarize. The phenomena contradicting common sense are (1) trace in a location without a trace that led to that location and (2) information being sent from one location to another without a trace of any information carriers. These phenomena are observed in our world. Hence, if one insists on a one-world interpretation of quantum mechanics, then one must concede that the laws of physics violate common sense explanation. We have argued that there is no need for such a concession. Common sense explanation can be restored by rejecting one-world interpretations and embracing the MWI. In the MWI, our world still contains the paradoxical phenomena of (1) and (2). But those phenomena arise out of, and can be locally explained in terms of, (multiple) continuous trajectories in the full physical universe that includes all worlds together. This is how the MWI brings common sense to paradoxical quantum experiments.

The MWI also brings common sense to other quantum experiments which, if considered in a single world, deny the possibility of common sense explanation. Bell type correlation experiments suggest "spooky action at a distance" (Vaidman 2015b). Quantum teleportation in a single world transfers huge amounts of information by sending just a few bits (Vaidman 1994).

It is widely believed that the MWI violates common sense and some researchers even reject the MWI on that basis. Our arguments challenge this sentiment. The parallel worlds of the MWI, when considered together, restore common sense causal explanation to physics. Perhaps one could still mount a common sense based objection to the MWI, if one thought that the unintuitive features of the MWI outweigh its intuitive causal explanations. This is a big issue: a proper analysis of all unintuitive aspects of the MWI requires a paper in itself. However, we will conclude with some brief reasons for thinking that the unintuitive aspects of the MWI are of little theoretical consequence.

There seem to be two main concerns. First, that we humans branch might seem unintuitive. Second, the sheer existence of parallel worlds might seem unintuitive. Regarding the first, it is well known that some organisms – amoeba – routinely branch and leave behind multiple descendants. This can seem surprising, until we understand the (biological) mechanism that gives rise to it (mitosis). We are organisms too, so provided we understand the (quantum) mechanism that gives rise to our branching (entanglement), then there should be no great reason for concern. Regarding the existence of parallel worlds, we can note that it was once deemed counterintuitive to entertain the thought of many parallel galaxies. Indeed, the astronomer Giordano Bruno was executed for making this suggestion. In hindsight, we now see that this hypothesis merely went against the widely held assumption that our solar system is special. Accordingly, there is no particular reason to consider our universe special by thinking it is the only one. And indeed, modern cosmology has gotten used to the idea of parallel worlds, for reasons independent of quantum mechanics, which have to do with cosmic inflation and the fine-tuning of the cosmological constants. Again, this is not to say that there is nothing unintuitive about the MWI. It is to say that in our view the unintuitive aspects are outweighed by what the MWI can offer common sense.

This work has been supported in part by the Israel Science Foundation Grant No. 2064/19 and the German–Israeli Foundation for Scientific Research and Development Grant No I-1275-303.14.